\documentclass[11pt]{article}
\title{Indigenizing the next decade of astronomy in Canada}
\usepackage{times}
\usepackage{geometry}
\usepackage[utf8]{inputenc}
\usepackage{enumitem,amssymb}
\usepackage{graphicx}

\usepackage[authoryear]{natbib}
\setcitestyle{authoryear,open={(},close={)}}
\usepackage{mdframed} 
\usepackage{fancyhdr}

\mdfdefinestyle{theoremstyle}{
innertopmargin=\topskip,}
\mdtheorem[style=theoremstyle]{lrptextbox}{}

\begin{document}
\begin{center}
{\Large{\bf Indigenizing the next decade of astronomy in Canada}} \\
Hilding R.~Neilson \footnote{Department of Astronomy \& Astrophysics, University of Toronto, Toronto, Ontario, M5S~3H4, Canada. \\ e-mail:hilding.neilson@utoronto.ca}, Laurie Rousseau-Nepton \footnote{Canada-France-Hawaii Telescope, Kamuela, HI, 96743, USA; Department of Physics and Astronomy, University of Hawaii at Hilo, Hilo, HI, 96720-4091, USA}, Samantha Lawler \footnote{Campion College, University of Regina, Regina, SK S4S 0A2, Canada} \& Kristine Spekkens \footnote{Department of Physics and Space Science, Royal Military College of Canada, Kingston, ON, K7K 7B4, Canada}
\end{center} 
\noindent{\bf Executive Summary} \\
The Truth and Reconciliation Commission of Canada published its calls to action in 2015 with 94 recommendations. Many of these 94 recommendations are directly related to education, language, and culture, some of which the Canadian Astronomy community can address and contribute to as part of reconciliation. The Canadian Astronomy community has an additional obligation since it benefits from facilities on Indigenous territories across Canada and the world. Furthermore, Indigenous people are still underrepresented at all levels in Canadian astronomy. The purpose of this Community Paper is to develop recommendations for the Canadian astronomy community to support Indigenous inclusion in the science community, support Indigenous learning by developing Indigenous-based learning materials and facilitate access to professionals and science activities, and to recognize and acknowledge the great contributions of Indigenous communities to our science activities.
As part of this work we propose the ten following recommendations for CASCA as an organization and throughout this Community Paper we will include additional recommendations for individuals: astronomers, students and academics.  The ten recommendations are:
\begin{enumerate}
\item That CASCA  partner with Indigenous organizations to support Indigenous education and diversity in astronomy and STEM.  CASCA should collaborate with Indigenous language teachers and programs to develop astronomical learning material in Indigenous languages. This could be done through a dedicated CASCA committee on Indigenizing Astronomy.
\item That CASCA collaborate with the National Research Council Herzberg, the Canadian Space Agency and other facilities to create programs in which a share of the telescope time is used for programs that promote education and inclusion for Indigenous and underrepresented people and schools in Canada and around the world. 
\item That CASCA expand the CASCA-Westar program, to fund training and support for volunteers and requires these programs to have multi-year plans to continually build and support connections with Indigenous communities and create long-lasting impact.
\item That CASCA create scholarships and funding specifically for Indigenous peoples to participate in physics and astronomy.
\item That CASCA advocate for diversity hiring initiatives at the postdoctoral and faculty level, and hold university policies to account. 
\item That CASCA should set targets to increase Indigenous participation at all levels of membership that are consistent with the general demographics and take actions to encourage greater involvement, including the creation of new membership categories.
\item Detailed historical, political, social and cultural context about Indigenous peoples in lands where Canada uses facilities such as Hawaii, Arizona, Chile, Australia, South Africa and other regions needs to be available to any CASCA members in leadership positions. This could be done by producing training material including information about Indigenous peoples in Canada and the different regions of the world where the Canadian astronomy community is active. 
\item That CASCA funds the development of a diversity training program for faculty, post-docs and students that directly educates on issues regarding Indigeneity in Canada and in the world.
\item That CASCA offers funding for students, post-docs and faculty to participate in diversity, equity, inclusion organizations such as .caISEs, SACNAS, that promote the inclusion of Indigenous peoples in STEM. 
\item That CASCA becomes an example for other countries for its programs regarding Indigenizing astronomy and share its educational material around the world.
\end{enumerate}
%


\noindent\textbf{Contributing Author bios}\\
Hilding Neilson is a contract-limited term appointment assistant professor in the Department of Astrophysics \& Astrophysics and an associate member of the Dunlap Institute for Astronomy \& Astrophysics at the University of Toronto.  He is Mi'kmaq First Nations and a member of the Qalipu First Nations band of the island of Ktaqmkuk (Newfoundland).  He is grateful to be a guest on the unceded territory of the Huron-Wendat, the Seneca and most recently the Mississaugas of the Credit River.   \\
\noindent Laurie Rousseau-Nepton is a member of the Pekuakamiulnuatsh First Nations community. She is the first woman from a First Nations community in Canada to get a PhD in Astronomy. She is a Canadian resident Astronomer working at the Canada-France-Hawaii Telescope on the Big Island of Hawaii. She wishes to recognize and acknowledge the very significant cultural role that the summit of Maunakea has always had within the indigenous Hawaiian community. She is most grateful to have the opportunity to conduct observations from this mountain. \\
\noindent Samantha Lawler is a first-year assistant professor at Campion College, University of Regina on the territories of the n\^ehiyawak, Anih\v{s}in\={a}p\={e}k, Dakota, Lakota, and Nakoda, and the homeland of the Metis.  She acknowledges her settler ancestry and is actively learning how to incorporate traditional Indigenous knowledge into her astronomy teaching.\\
Kristine Spekkens is an associate professor at the Royal Military College of Canada who is cross-appointed at Queen's University, which is situated on traditional Anishinaabe and Haudenosaunee territory. She is grateful to be able to live and learn on those lands. She acknowledges her settler privilege and is learning how to support Indigenous inclusion within her research, teaching and service to the astronomical community.

\section{Introduction}
The night sky has been studied and used by people from all over the world to remember important moments, transfer knowledge, inspire, connect to each other, travel, survive and understand the universe, yet this reality is not apparent when we consider astronomy textbooks, courses, and physics/astronomy departments.  This is particularly concerning in Canada as we live on Indigenous territories from coast to coast to coast and is also concerning as Canadian astronomy benefits from access to Indigenous territories around the world. Most astronomy textbooks have only a few pages, if any, dedicated to astronomy knowledge from Indigenous cultures and most of those pages references Indigenous peoples and Indigenous knowledges in the past tense, as if Indigenous peoples are extinct or irrelevant.  The astronomy knowledge that does get presented tends to be superficial, tends to describe Indigenous peoples from an anthropological perspective and presents knowledge as religious and ceremonial. Rarely is Indigenous knowledge presented as scientific.  This  erasure of both peoples and knowledges  is problematic because it excludes Indigenous people from discussion in astronomy and from seeing themselves and their knowledge reflected in the modern days astronomy in Canada \citep{cajete2000native, battiste2017decolonizing}.

This lack of consideration is reflected in the demographics of CASCA and in the number of Indigenous scholars in the astronomy community and in STEM, in general.  There is  not a complete census of the demographics in STEM fields  in Canada, however, we can consider the 2017 National Science Foundation report as analogous\footnote{https://www.nsf.gov/statistics/2017/nsf17310/}.  For most fields discussed in that report the number of Indigenous PhDs in STEM is written as too few to be statistical.  This is likely the same story in Canada where if PhD graduates were representative of the broader population about 1 PhD in every 20 would be earned by an Indigenous person.  

In a similar vein Indigenous peoples in Canada statistically have poorer education outcomes at all stages of education\footnote{https://www150.statcan.gc.ca/n1/daily-quotidien/171025/dq171025a-eng.htm?indid=14430-1\&indgeo=0}. These outcomes occur for numerous reasons relating to life on reserves, income inequality, multi-generational traumas related to ongoing colonization, and so on. Therefore, resolving the outcome differences between Indigenous students and non-Indigenous students is of importance and will require input from scientists of all fields, not just astronomers.

There are a number of key issues in the Canadian astronomy community can address:

\begin{enumerate}
\item the lack of education (courses, training, etc.) regarding historical, political, social and cultural contexts regarding Indigenous peoples in Canada that limits the ability for CASCA members to understand our roles and responsibilities to Indigenize science and astronomy and to understand our place on this land; 
\item the dearth of Indigenous students and scholars in the Canadian community;
\item the lack of educational resources and outreach that are devoted to Indigenous communities;
\item the lack of interaction and collaboration with Indigenous communities;
\item the need to diversify our definitions within science and of science that are exclusive of Indigenous knowledges.
\end{enumerate}

To address the current inequities that exist in Canadian astronomy, we present a discussion of Indigeneity in Canada and the role of CASCA, as an organization that benefits from the use of ceded and unceded territories and from ongoing colonization. We present this discussion in the context of Equity, Diversity and Inclusion, the context of education and outreach, and in the context of enabling researchers from all horizons to be a part of  scientific discovery in the next decade.  It is the aspiration of the authors that CASCA and the Canadian astronomy community in general will offer leadership to create an academic environment that encourages and supports Indigenous people in the study of astronomy and astrophysics. 

In an ideal future, our astronomy courses would share an excitement of discovery and learning that braids Indigenous knowledges and Western/Eurocentric methods \citep[e.g.,][]{hatcher2009two,kimmerer2013braiding}.  In this future our curricula, meetings and interactions would engage with scientific progress and Indigenous  cultures and methods collaboratively and not engage simply as separate boxes-to-check.  It is expected that there will be a growing population of Indigenous scholars at all levels of education who work in both Indigenous and non-Indigenous communities. As such we will be developing connections and continual interactions with Elders, knowledge keepers and teachers in those communities stretching from coast-to-coast-to-coast that acts to bring education and outreach to both Urban Indigenous youth and Indigenous youth in under-served regions. As we develop these connections and interactions, a better future would mean we have created a path that encourages and appreciates Indigenous peoples to engage in astronomy and STEM.  It would further include a mutual and constructive respect of Indigenous knowledges and methodologies in concert with Western science.

Throughout this paper we discuss the process of Indigenizing astronomy as opposed to the more common phrase of decolonization.  \cite{tuck2012decolonization} offer a detailed discussion of when and how it is appropriate to discuss decolonization.  Decolonization should not employed to describe metaphorical concepts such as decolonizing astronomy.  Therefore, we choose not to use this word.  It is notable that the descriptor ``Indigenizing'' is also not ideal.   The act of Indigenizing a field is not one that can be led by non-indigenous organizations or groups.  As such, in principle CASCA cannot Indigenize astronomy, but it can be a leader in creating space for Indigenous peoples and knowledges.  Finally, we avoid the term ``White Paper''.  This term has historical significance in Canada based on a number of white papers produced by federal governments in the 1960s designed to remove rights of Indigenous peoples in this country \citep{weaver1981making}.  As such we refer to this as a Community Paper.  

The goal of this Community Paper is to describe our recommendations and offer ideas for how astronomers and astronomy organizations can improve the representation of Indigenous peoples and knowledges in our field.  In the next section we provide a brief overview of what it means to be Indigenous in Canada in terms of both history and identity.  This is followed by a summary of the Truth and Reconciliation Calls to Actions and those calls that directly influence the recommendations of this community paper. In Section 4, we note the current challenges for Indigeneity in astronomy in terms of education, hiring, and outreach, along with discussion of how we as Canadians be better aware of our role and responsibilities on these lands. We follow this with discussion of embracing Indigenous knowledges in the classroom and in research, and how CASCA can support that action.  In section 5, we expand the concepts of embracing knowledges in astronomy with discussion of how we can grow connections with Indigenous communities and discuss issues of cultural sensitivity in Section 6.  We conclude with a summary of the recommendations and discussion of how these could be prioritized by the astronomy community as a whole.

\section{Historical Background and Calls for Action}
Indigenous peoples have lived across the land that we now call Canada for tens of thousands of years.  Many Indigenous peoples tell creation stories that connect to the land since time immemorial. According to recent Statistics Canada census; Indigenous people make up about 5\% of the population in Canada.  The government of Canada recognized Indigenous peoples or the first peoples of Canada in three categories differentiated among other things by their governance structure: First Nations, Inuit, and M\'etis.  

Inuit peoples are from territories in northern Canada in four primary regions: Nunavut; the Inuvialuit Settlement Region (Northwest Territories); Nunavik (Quebec); Nunatsiavut (Labrador)\footnote{https://www.itk.ca/about-canadian-inuit/}. The Inuit people also live in Alaska, Greenland, and Siberia.  According to Statistics Canada, in 2016 there were about 65,000 Inuit people in Canada with an average age of 27.7 years, more than a decade younger on average than the general population of Canada.

The M\'etis people have a unique culture as a mixture of European and First people heritage, who settled across Canada and the US from what is now Ontario to eastern British Columbia.  According to Statistics Canada there were about 600,000 people identifying as M\'etis in 2016.  M\'etis peoples were only recognized as Indigenous by the Canadian government in 1982. Currently they do not have recognized land rights like the ones associated with First Nations and Inuit peoples.  

The First Nations peoples in Canada refers to peoples from across Turtle Island that includes more that 600 communities, from the Mi'kmaq peoples in the East to the Salish in the West, to name a few.  In Canada there are more than 600 First Nations communities with more than fifty nations with distinct languages.  According to Statistics Canada, almost one million people in Canada identify as First Nation.  The average age of First Nations peoples is about thirty years old, much younger than the population in general. It should be noted that only about 260,000 First Nations people identify as speaking a First Nations language.

Of the First Nations peoples, the Canadian government separates identity into two parts: status and non-status.  Status is an identity recognized by the federal government as part of the Indian Act from 1867.  There are a number of historical issues with this classification \citep[see][for example]{joseph201821} that relate to assimilation and removal of identity.   Today, status is related to identifying with a federally recognized First Nations band and is meant to share education and health-related benefits.  As of the 2016 census about three-quarters of First Nations people have status.  Furthermore, less than one-half of First Nations peoples live on federally recognized Reserves.

Today many Indigenous peoples live in urban areas.  According to the 2016 Census, there are almost 100,000 Indigenous people living in Winnipeg, followed by Edmonton, Vancouver and Toronto  having the largest urban Indigenous populations.  This means that Indigenous peoples and cultures need to be recognized as part of the urban fabric and that we, in universities, have an obligation to be aware of the cultural differences and similarities of First Nations, Inuit, and M\'etis peoples from across Turtle Island and the world.

Indigenous people have been on Turtle Island since time immemorial. The advent of colonization from Columbus, Cabot, Cartier, and others have had negative impacts on the Indigenous peoples.  For instance, the Mi'kmaw populations of eastern North America have been devastated by war with the English, spread of diseases, displacement by colonization, and scalping bounties; all before confederation of Canada.   
With confederation came the Indian Act, whose goal was to eliminate ``the Indian problem''. The Indian Act continued the reserve system that removed many Indigenous peoples from their traditional territories in order to enable commercial and industrial development on the land without interference. It also created the Residential School system.  The philosophy of the residential school system was to ``save the child, kill the Indian'' and was designed to remove Indigenous children from their homes to be taught in boarding schools where their identities and cultures were banned.  Children at these schools were typically treated poorly with minimum nutrition and support. Many children experienced physical, mental, and sexual abuse. These schools contributed to multi-generational traumas that resonate today (www.trc.ca), traumas that have exacerbated in many ways the inequalities between the First peoples of Canada and the other Canadians.

Along with the residential schools, the Canadian government enacted numerous policies designed to disenfranchise Indigenous peoples including, but not exclusive to: Indian passes to leave reserves, removal of Indian status to attend university, removal of Indian status to serve in the military, removal of Indian status for Indigenous women marrying non-Indigenous men, etc. The reserve system was designed to contain Indigenous peoples on parcels of land much smaller than their traditional territories.  Not only were Indigenous people limited to the reserves, but if they wanted to travel elsewhere, they required a pass from the local Indian agent, who was a representative of the federal government. Along with this if Indigenous people wanted to attend university, they would have to give up status and then might not be able to return to reserves.  It was a choice between an education and family and still today First Nations people are reluctant to send their kids to study outside from the reserves since most do not come back after graduation. 

These acts of continued colonization contributed to the erasure of Indigenous knowledges and cultures.  Canada has begun to recognize the history and the current practises of colonization through a number of commissions and reports: from the Royal Commission on Aboriginal Peoples in 1996, the Truth and Reconciliation Commission in 2015, and the Commission for Missing and Murdered Indigenous Women in 2019.  The results for these commissions have argued that the treatment of Indigenous peoples by Canada have amounted to cultural genocide \citep{truth2015final} and to genocide (Reclaiming Power and Place: The Final Report of the National Inquiry into Missing and Murdered Indigenous Women and Girls; http://www.mmiwg-ffada.ca/).  As part of these commissions, there have been many recommendations - we focus here on the Truth and Reconciliation Commission Calls to Action that included 94 recommendations (www.trc.ca).  

The Truth and Reconciliation Commission of Canada published its calls to action in 2015 with 94 recommendations. Many of these 94 recommendations are directly related to education, language, and culture, some of which the Canadian Astronomy community can address and contribute to as part of the reconciliation process. The Canadian Astronomy community has an additional obligation since it benefits from the presence of facilities on Indigenous territories across Canada and the world and that Indigenous people are still underrepresented at all levels in Canadian astronomy. The purpose of this community Paper is to develop recommendations for the Canadian astronomy community to support Indigenous learning, support Indigenous inclusion, develop Indigenous-based learning materials for the classroom, and to recognize our role and responsibility toward the reconciliation with Indigenous people of Canada and the world.

\section{Indigeneity and Astronomy in Canada}
Embracing Indigenous knowledges and addressing ongoing inclusion issues in astronomy are crucial if our field is ever going to be as productive as it can be.  At the time of this work, there are almost no Indigenous peoples in academic astronomy and physics in Canada.  Furthermore, we cannot provide demographic numbers because CASCA has historically not tracked them and neither has NSERC.  If we assume the demographics in physics and astronomy in Canada are similar to that in the United States, then following the NSF survey there are very few Indigenous people with doctorates in physics and astronomy. For many categories in the NSF survey, it is claimed there are too few Indigenous scholars to count, that is there are statistically zero scholars in that category.  The problematic dearth of Indigenous scholars in science was unintentionally highlighted in an article by the Washington Post\footnote{https://www.washingtonpost.com/local/education/the-science-divide-why-do-latino-and-black-students-leave-stem-majors-at-higher-rates/2019/05/03/e386d318-4b32-11e9-93d0-64dbcf38ba41\_story.html}. In that article the author noted that Black and Latinx scholars are underrepresented in the sciences, but did not mention Indigenous scholars at all.  The article erased Indigenous people from the discussion.  The Washington Post article shows how Indigenous people and Indigenous knowledges have been ignored in the greater conversation around diversity and inclusion. This is not a unique example. 

While looking at how we learn and teach astronomy and physics and the underlying basis for that teaching, we can note that most astronomy textbooks and courses view knowledge from a western lens and dismiss non-western knowledges as beliefs or religion. This is largely a consequence of colonization that acts to dismiss and erase Indigenous knowledges and that ranks European knowledges as more important.  This way of learning and research is present not only in Astronomy but in all science fields.  From an Indigenous perspective, we can only imagine how frustrating this way of teaching can be and how excluded they can feel.

A first step to connect with Indigeneous knowledges and teachings is to listen to Indigenous Elders and teachers and  to learn about their ways, cultures and peoples.  CASCA has an opportunity and an ethical duty to help build pathways for Astronomers and teachers to connect and collaborate with Indigenous teachers and knowledge keepers. One pathway would be offering equity and diversity training for CASCA board members and the membership in general and engaging with Indigenous advisors for decision making.  CASCA should also advocate that universities continue to improve hiring practices and to promote equity training for hiring committees.  Finally, CASCA should set targets to increase Indigenous participation at all levels of membership that are consistent with the general demographics and take actions to encourage greater involvement.

\section{Role of Indigenous knowledges in the classroom}
Part of improving inclusion in astronomy includes embracing Indigenous knowledges and methodologies in the classroom.  Astronomy textbooks typically consider Indigenous knowledges when discussing constellations or solar/lunar calendar.  The discussion tends to treat Indigenous knowledge as ancient history and as something less valuable than western or Eurocentric science.  This erases Indigenous knowledges in our classes and can exclude Indigenous students. By simply changing the context in which Indigenous astronomical knowledge is presented and how it is presented, we can be more inclusive and culturally sensitive.  This has to be done in collaboration with Indigenous communities and training material should be provided for that specific purpose. 

There are many reasons that could explain erasure including racism and colonization. Another reason is our dependence on the axiom of the scientific method and falsifiability. The scientific method largely arose during the Enlightenment in Europe and is largely credited to Sir Francis Bacon \citep{shapin2018scientific}.  This scientific method required that science grow based on direct experiment and experience.  This shift in knowledge development replaced knowledge systems in Europe.  In the 20th century, Karl Popper codified the scientific method as a method of falsification \citep{popper1963science}.  Today, the astronomy community tends to ignore knowledges that are not seen as fitting within that scientific methodology.  However, embracing Indigenous knowledges as part of astronomy would benefit our community by being more inclusive and by being more respectful of where we live.  Furthermore, doing this would benefit our science.  \cite{lipe201919} identified a number of significant differences between Western science and Indigenous knowledge systems.  Some of the more relevant to astronomy include: 
\begin{enumerate}
\item Western science focuses on average behaviours of phenomena, Indigenous knowledge on extreme events;
\item Western science considers us as independent of and apart from phenomena, Indigenous knowledge considers phenomena as relational and connected to us;
\item Western science tends to isolate phenomena; Indigenous knowledge considers phenomena in a wholistic sense;
\item Relative to Indigenous knowledges, Western science considers information only over relatively short time spans, Indigenous knowledges contains information spanning much longer timescales.
\end{enumerate}
These differences offer complementary methods to understand the Universe.  For example \cite{Leaman:1692550} noted that in the Australian Aboriginal story of Nyeeruna there is knowledge of the light variability of the star Betelguese. The Aboriginal peoples ``discoverered'' Betelguese was a variable stars centuries before Western science and observed variability of other red giant stars well before Western science \citep{hamacher2018observations}.

Considering Indigenous methods offers new directions in astronomy teaching that change how students work with data and understand the Universe.  We recommend that astronomy teachers work to consider Indigenous knowledges in their astronomy classes and be better prepared to present the information in a more sensitive way.   For instance, one new professor (SML) has begun the very first steps to integrate Indigenous knowledge into an Astronomy 101 class that did not previously include any. The course is offered through Campion College, an affiliated college of the University of Regina (U~of~R). The U~of~R has a large undergraduate Indigenous population, about 12.7\% (https://www.uregina.ca/profile/). The U~of~R has an Office of Indigenization, and one of their stated goals is to help connect faculty and staff with Indigenous Elders/knowledge keepers and to help with indigenization of their courses. The new professor was able to make use of their Elder-in-residence program to bring a Life Speaker to the first class. The staff of the Office of Indigenization worked with the professor to help with pronunciations to properly acknowledge the territory, and with the proper etiquette for introducing the Life Speaker to the class, as well as providing her with a properly prepared traditional offering of tobacco to present to the Life Speaker prior to him speaking to the class. The Life Speaker was asked to speak about his traditional knowledge of astronomy, and he spoke at length about the traditional view of the Earth as our mother and the Universe as our father, as well as the personal connection with the Northern Lights. He used the Cree name for the Northern Lights, and translated it to ``the relatives are dancing.''

Starting the course with an acknowledgement of the land and a Life Speaker's very personal, interconnected view of astronomy completely changed the starting point of the class, and created a much more respectful beginning to this course where the standard is often to start with ``primitive'' views that were later shown to be ``false.'' This started the course on a much more respectful note, allowing for traditional knowledge to coexist with current scientific practices as another way of knowing. 

With extensive help from the U~of~R Office of Indigenization, a brand-new professor was able to properly acknowledge a land that was new to her, and bring in the traditional knowledge of an Indigenous Elder, on her very first day of teaching. Administrative procedures were already in effect to make compensating the elder for his time with a proper honorarium from Campion College, which is also an important part of acknowledging the importance of traditional knowledge. Having the Office of Indigenization as a resource for advice and for connection with indigenous groups is completely invaluable for indigenizing astronomy (as well as other courses).
SML is extremely thankful that these structures exist and acknowledges that without the help of the Office of Indigenization, it would have taken years for her to build up the contacts to do what she was able to present on her very first day of teaching. In the near future, she looks forward to expanding collaboration with their office and with First Nations University, another affiliated college of the University of Regina.

The simple step of connecting educators with Indigenous Elders impacts how we deliver astronomy content. We ask that CASCA work with and support organizations and groups that work with Indigenous knowledges.  Even simply offering support to bring Indigenous Elders and/or knowledge keepers to classrooms can positively impact how we approach Indigenous knowledges. We recommend that CASCA provide resources to promote Indigenization of astronomy. This could be done by forming a new committee with a precise mandate to  lead the effort to indigenize astronomy.

\section{Reaching Indigenous Communities}
While we need to be more inclusive of Indigenous knowledges in our classrooms, we also need to work to build relationships with Indigenous communities in Canada and with Indigenous communities where astronomy facilities are located.  CASCA has been promoting the Westar and the Astronomy for Canadian Indigenous People (ACIP) (Moumen et al 2019) programs as such avenues for nourishing these relationships. 

The CASCA Westar lecture program helps connect underserved communities to Canadian Astronomers, organize public lectures and gives volunteers the opportunity to engage and learn about different views of the night sky while they visit the communities. The ACIP program gives public lectures and school presentations in specific reserves (different every year) and organizes and funds a two-day event where Indigenous students and some chaperones from their community visit the Mont-M\'{e}gantic observatory and experience a night of observation with an astronomer. 
While these programs have had successes they are not reaching enough communities to have a significant impact given the more than six hundred Indigenous communities and more than three thousand reserves in Canada as well as the increasing populations of urban Indigenous peoples.  Also, visiting communities once or twice does not contribute to building  relationships or earning the needed trust for astronomy to become more inclusive.  Nurturing these relationships requires continued engagements over numerous years.  Nevertheless, these initiatives offers a good starting point to develop a more inclusive program. 
Nurturing these relationships requires the astronomy community to be welcoming of Indigenous Elders, Indigenous knowledge keepers and Indigenous teachers. Recent annual meetings of CASCA have featured invited talks from Elders.  This is a promising beginning for connecting with communities.  We can improve this by inviting more knowledge keepers to meetings and by opening CASCA membership to peoples in Indigenous communities. 

In summary, we recommend that the ACIP and Westar programs be expanded with additional funding and the volunteers be encouraged to provide multi-year plans for repeated engagements along with facilitated access to training material for them. This would allow us to nurture relationships and move past the perception of parachuting people into communities to ``educate'' Indigenous peoples.  We also recommend funding and subsidizing membership and participation in CASCA for Indigenous educators beyond the standard corporate versus academic membership standards.

\section{Avoiding Appropriation, Racism and Patronizing Attitudes while Improving Inclusion}
The academic culture in Canada and around the world has been traditionally unwelcoming to Indigenous learners, Indigenous teachers, and Elders.  There has been a history of appropriating Indigenous knowledges and ignoring Indigenous voices. Shifting from this culture to one that is inclusive of Indigenous communities, learners and knowledges requires us to collaborate with Indigenous communities and to listen to Indigenous voices. This shift requires the Canadian astronomical community to be considerate of the feelings, interests and needs of the Indigenous communities and for us to connect with Indigenous knowledges in the ways communities require. It is a sensitive matter and Academia needs to adapt to these changes while respecting the differences.

As part of the response to calls to action of the Truth and Reconciliation Commission, many universities have responded by creating programs for hiring Indigenous scholars.  This is a productive step at inclusion in academia, but is a step that requires departments to commit to Indigenization.  One author, HRN, was invited to interview for one such position. This was done because he self-identified as Indigenous as part of a different faculty search.  During the interview, faculty at that institution questioned his identity, asking ``How Indigenous are you?''.  Faculty also claimed during the interview that Indigenous knowledge was irrelevant and not scientific, after HRN gave an education talk about Indigenizing astronomy and the need to respect Indigenous ways of knowing.  This response was inappropriate and led to complaints from the Indigenous Elder involved in the interview process and by HRN.   Although it might seem trivial to one, such questions as ``How Indigenous are you?'' are inappropriate during the hiring process and hiring committees should be trained to avoid such situations and be aware of the unconscious biases that negatively impact hiring candidates from underrepresented groups. 

This specific experience relates to two issues with Indigenizing astronomy in Canada - the first being tokenism,  and the other being how we, as a community, treat Indigenous knowledges.  Because there is institutional pressure to hire Indigenous scholars there is the risk of persons falsely self-identifying as Indigenous or identifies as such because a distant relation was Indigenous.  In Canada, Indigenous identity is determined by communities and is related to how connected a person is to the community, not just blood line or vague ancestry or DNA tests.  This may have been the motivation when HRN was asked how Indigenous he is by the (non-Indigenous) faculty member.  Because HRN is also white, the person may have been concerned that there was appropriation of Indigenous identity. Training could have prevented this situation to happen both by improving the understanding of the Indigenous identity in Canada and the difference and validity of Indigenous knowledge-based information.  

The second issue is how we treat Indigenous knowledges.  When faculty members dismiss Indigenous knowledges, they  dismiss millennia of experiences, history and identity.  Doing so during a faculty interview was discrimination and highlights a clear lack of understanding or appreciation of Indigenous knowledges for astronomy. This also highlights an additional issue of the bystander effect.  These incidents were viewed by numerous other faculty who remained silent. At no point did anyone intervene.  Even after that interview was completed and the incident become known more broadly in the institution there was still no meaningful response from faculty in that institution. This experience demonstrates a need for training about Indigenous knowledges and working with Indigenous peoples. There is also a need for training with respect to discrimination and the bystander effect.

Such training is important since it can have great impacts on how Canadians perceive the Indigenous peoples in Canada and understand their role and responsibility in the reconciliation process.  Every person as the right to feel like they belong and to feel that their knowledges are respected.  We as an astronomy community have an ethical duty to live up to that responsibility.

\section{Visions for 2030}
What could astronomy in Canada look like if we Indigenized today? If we put in the effort today, we will see an increase in the number of students from Indigenous communities that choose science-related disciplines for their career. When these students reach graduate school programs and eventually become post-doctorate and faculty members, our organization will benefit from the more diverse and inclusive environment it will have created and from the discoveries that will be made by these researchers.  In order to reach a point where our Canadian astronomy community is representative of the Canadian population (about 5\% of membership), we have to start measuring now. This includes surveying the demographic profiles of the members of CASCA and more broadly the student population in Physics and Astronomy. We need to set ambitious goals to increase Indigenous representation at the undergraduate level if there is to be a continuing cohort of Indigenous graduate students, as opposed to one or two per generation that appears to be current reality. 

If we put the effort today, we will nurture better connections and relationships with communities across Canada. We will provide classrooms with learning material that is appropriate, considerate and inclusive so they can see Indigeneity in nature and in practice. Having Indigenous astronomy education material available at all levels will help retain  Indigenous students in the field and be inclusive of the next generation of researchers to come.   There are too few Indigenous tenure-stream faculty in astronomy and physics currently. Increasing Indigenous representation in the faculty population is a crucial step forward since it would reflect a strong desire for change and inclusion.  If we put in the effort today, we hope that by the next Long Range Plan we will have a professional culture of understanding that values Indigenous knowledges and understands our role and responsibility on Turtle Island.






\begin{lrptextbox}[How does the proposed initiative result in fundamental or transformational advances in our understanding of the Universe?] 
This initiative acts to create a more inclusive field for astronomy by embracing Indigenous knowledges and work with the communities. In will result in fundamental advances in our understanding of the different communities across Canada and the Universe. A more inclusive environment can only help with our productivity in the field.  
\end{lrptextbox}

\begin{lrptextbox}[What are the main scientific risks and how will they be mitigated?]
N/A
\end{lrptextbox}

\begin{lrptextbox}[Is there the expectation of and capacity for Canadian scientific, technical or strategic leadership?] 
N/A
\end{lrptextbox}

\begin{lrptextbox}[Is there support from, involvement from, and coordination within the relevant Canadian community and more broadly?] 
Indigenizing astronomy and STEM relates to the goals of NSERC and the Canada Research Chair programs in terms of equity, diversity and inclusion.
\end{lrptextbox}

\begin{lrptextbox}[Will this program position Canadian astronomy for future opportunities and returns in 2020-2030 or beyond 2030?] 
Yes. By being more inclusive and open to differences, Canadian will be in better position for collaborations all around the world. 
\end{lrptextbox}

\begin{lrptextbox}[In what ways is the cost-benefit ratio, including existing investments and future operating costs, favourable?] 
N/A
\end{lrptextbox}

\begin{lrptextbox}[What are the main programmatic risks
and how will they be mitigated?] 
N/A
\end{lrptextbox}

\begin{lrptextbox}[Does the proposed initiative offer specific tangible benefits to Canadians, including but not limited to interdisciplinary research, industry opportunities, HQP training,
EDI,
outreach or education?] 
The initiative acts to benefit EDI; outreach and education; interdisciplinary and transdisciplinary research directly.
\end{lrptextbox}

\bibliography{WPIA-refs}

\end{document}